\documentclass{article}

\usepackage[preprint,nonatbib]{neurips_2022}

\usepackage[utf8]{inputenc} %
\usepackage[T1]{fontenc}    %
\usepackage{hyperref}       %
\usepackage{url}            %
\usepackage{booktabs}       %
\usepackage{mathtools}
\usepackage{amsfonts,amssymb}       %
\usepackage{nicefrac}       %
\usepackage{microtype}      %
\usepackage{xcolor}         %

\usepackage[ruled, vlined, linesnumbered]{algorithm2e}
\usepackage{algorithmic}
\usepackage{braket}
\usepackage[qm]{qcircuit}
\usepackage{graphicx}

\usepackage{bbm}

\usepackage{multirow}

\definecolor{qcmrfcol}{RGB}{0,0,0}

\usepackage{tikz}
\tikzset{%
pics/model/.style 2 args={code={
\ifnum#1=0
\draw [qcmrfcol,very thick] (0.25,-0.25) circle (0.5mm);
\fi
\ifnum#1=1
\draw [qcmrfcol,very thick] (0,-0.25) circle (0.5mm);
\draw [qcmrfcol,very thick] (0.5,-0.25) circle (0.5mm);
\draw [qcmrfcol,thick] (0.03,-0.25) -- (0.47,-0.25); %
\fi
\ifnum#1=2
\draw [qcmrfcol,very thick] (0,0) circle (0.5mm);
\draw [qcmrfcol,very thick] (0.5,0) circle (0.5mm);
\draw [qcmrfcol,very thick] (0,-0.5) circle (0.5mm);
\draw [qcmrfcol,thick] (0.03,0) -- (0.47,0); %
\draw [qcmrfcol,thick] (0,-0.47) -- (0,-0.03); %
\fi
\ifnum#1=3
\draw [qcmrfcol,very thick] (0,0) circle (0.5mm);
\draw [qcmrfcol,very thick] (0.5,0) circle (0.5mm);
\draw [qcmrfcol,very thick] (0.5,-0.5) circle (0.5mm);
\draw [qcmrfcol,very thick] (0,-0.5) circle (0.5mm);
\draw [qcmrfcol,thick] (0.03,0) -- (0.47,0); %
\draw [qcmrfcol,thick] (0.03,-0.5) -- (0.47,-0.5); %
\draw [qcmrfcol,thick] (0,-0.47) -- (0,-0.03); %
\fi
\ifnum#1=4
\draw [qcmrfcol,very thick] (0,0) circle (0.5mm);
\draw [qcmrfcol,very thick] (0.5,0) circle (0.5mm);
\draw [qcmrfcol,very thick] (0.5,-0.5) circle (0.5mm);
\draw [qcmrfcol,very thick] (1,-0.5) circle (0.5mm);
\draw [qcmrfcol,very thick] (0,-0.5) circle (0.5mm);
\draw [qcmrfcol,thick] (0.03,0) -- (0.47,0); %
\draw [qcmrfcol,thick] (0.03,-0.5) -- (0.47,-0.5); %
\draw [qcmrfcol,thick] (0.53,-0.5) -- (0.97,-0.5); %
\draw [qcmrfcol,thick] (0,-0.47) -- (0,-0.03); %
\fi
\ifnum#1=5
\draw [qcmrfcol,very thick] (0,0) circle (0.5mm);
\draw [qcmrfcol,very thick] (0.5,0) circle (0.5mm);
\draw [qcmrfcol,very thick] (0.5,-0.5) circle (0.5mm);
\draw [qcmrfcol,very thick] (0,-0.5) circle (0.5mm);
\draw [qcmrfcol,thick] (0.03,0) -- (0.47,0); %
\draw [qcmrfcol,thick] (0.5,-0.47) -- (0.5,-0.03); %
\draw [qcmrfcol,thick] (0.03,-0.5) -- (0.47,-0.5); %
\draw [qcmrfcol,thick] (0,-0.47) -- (0,-0.03); %
\fi
\ifnum#1=6
\draw [qcmrfcol,very thick] (0.25,0.0) circle (0.5mm);
\draw [qcmrfcol,very thick] (0.5,-0.5) circle (0.5mm);
\draw [qcmrfcol,very thick] (0,-0.5) circle (0.5mm);
\draw [qcmrfcol,thick] (0.03,-0.5) -- (0.47,-0.5); %
\draw [qcmrfcol,thick] (0.2288,-0.021213) -- (0.021213,-0.4788); %
\draw [qcmrfcol,thick] (0.2712,-0.021213) -- (0.4788,-0.4788); %
\fi
\ifnum#1=7
\draw [qcmrfcol,very thick] (0.25,0.0) circle (0.5mm);
\draw [qcmrfcol,very thick] (0.5,-0.5) circle (0.5mm);
\draw [qcmrfcol,very thick] (0,-0.5) circle (0.5mm);
\draw [qcmrfcol,very thick] (0.75,0.0) circle (0.5mm);
\draw [qcmrfcol,very thick] (1,-0.5) circle (0.5mm);
\draw [qcmrfcol,thick] (0.03,-0.5) -- (0.47,-0.5); %
\draw [qcmrfcol,thick] (0.2288,-0.021213) -- (0.021213,-0.4788); %
\draw [qcmrfcol,thick] (0.2712,-0.021213) -- (0.4788,-0.4788); %
\draw [qcmrfcol,thick] (0.53,-0.5) -- (0.97,-0.5); %
\draw [qcmrfcol,thick] (0.7288,-0.021213) -- (0.521213,-0.4788); %
\draw [qcmrfcol,thick] (0.7712,-0.021213) -- (0.9788,-0.4788); %
\fi
\ifnum#1=8
\draw [qcmrfcol,very thick] (0,0) circle (0.5mm);
\draw [qcmrfcol,very thick] (0.5,0) circle (0.5mm);
\draw [qcmrfcol,very thick] (0.5,-0.5) circle (0.5mm);
\draw [qcmrfcol,very thick] (0,-0.5) circle (0.5mm);
\draw [qcmrfcol,thick] (0.03,0) -- (0.47,0); %
\draw [qcmrfcol,thick] (0.5,-0.47) -- (0.5,-0.03); %
\draw [qcmrfcol,thick] (0.03,-0.5) -- (0.47,-0.5); %
\draw [qcmrfcol,thick] (0,-0.47) -- (0,-0.03); %
\draw [qcmrfcol,thick] (0.4788,-0.021213) -- (0.021213,-0.4788); %
\draw [qcmrfcol,thick] (0.021213,-0.021213) -- (0.4788,-0.4788); %
\fi
}}}

\newcommand{\model}[2]{\scalebox{#1}{
\begin{tikzpicture}
\path (0,0) pic {model=#2};
\end{tikzpicture}
}}

\newcommand{\mR}{\mathbb{R}}
\newcommand{\mU}{\mathbb{U}}

\newcommand{\mC}{\mathbb{C}}
\newcommand{\mP}{\mathbb{P}}
\newcommand{\mQ}{\mathbb{Q}}

\newcommand{\cX}{\mathcal{X}}

\newcommand{\cH}{\mathcal{H}}

\newcommand{\cC}{\mathcal{C}}

\newcommand{\cS}{\mathcal{S}}

\newcommand{\cO}{\mathcal{O}}

\newcommand{\cD}{\mathcal{D}}

\newcommand{\bt}{\boldsymbol{\theta}}
\newcommand{\bbt}{\bar{\bt}}
\newcommand{\bgam}{\boldsymbol{\gamma}}

\newcommand{\br}{{\boldsymbol{r}}}

\newcommand{\bx}{\boldsymbol{x}}

\newcommand{\bbo}{\boldsymbol{1}}

\newcommand{\ba}{\boldsymbol{a}}

\newcommand{\bC}{\boldsymbol{C}}

\newcommand{\bmu}{\boldsymbol{\mu}}

\newcommand{\by}{\boldsymbol{y}}

\newcommand{\bX}{\boldsymbol{X}}

\newcommand{\otrace}{\operatorname{Tr}}

\newcommand{\In}{I^{\otimes n}}

\newtheorem{theorem}{Theorem}[section]

\newtheorem{corollary}[theorem]{Corollary}
\newtheorem{definition}[theorem]{Definition}

\title{On Quantum Circuits for Discrete Graphical Models}

\author{%
  Nico Piatkowski \\
    Fraunhofer IAIS\\
    ME Group\\
    53757 Sankt Augustin, Germany\\
  \texttt{nico.piatkowski@iais.fraunhofer.de} \\
  \And
  Christa Zoufal \\
IBM Quantum\\
IBM Research -- Zurich\\
8803 Rüschlikon, Switzerland\\
 \texttt{ouf@zurich.ibm.com} \\
}

\begin{document}

\maketitle

\begin{abstract}
Graphical models are useful tools for describing structured high-dimensional probability distributions. Development of efficient algorithms for generating unbiased and independent samples from graphical models remains an active research topic. Sampling from graphical models that describe the statistics of discrete variables is a particularly challenging problem, which is intractable in the presence of high dimensions. In this work, we provide the first method that allows one to provably generate unbiased and independent samples from general discrete factor models with a quantum circuit. Our method is compatible with multi-body interactions and its success probability does not depend on the number of variables. To this end, we identify a novel embedding of the graphical model into unitary operators and provide rigorous guarantees on the resulting quantum state. Moreover, we prove a unitary Hammersley-Clifford theorem---showing that our quantum embedding factorizes over the cliques of the underlying conditional independence structure. Importantly, the quantum embedding allows for maximum likelihood learning as well as maximum a posteriori state approximation via state-of-the-art hybrid quantum-classical methods. Finally, the proposed quantum method can be implemented on current quantum processors. Experiments with quantum simulation as well as actual quantum hardware show that our method can carry out sampling and parameter learning on quantum computers.
\end{abstract}

\section{Introduction}
Modelling the structure of direct interaction between distinct random variables is a fundamental sub-task in various applications of artificial intelligence \cite{Wang/etal/2013a}, including natural language processing \cite{Lafferty/etal/2001a}, computational biology \cite{Kamisetty/etal/2008a}, sensor networks \cite{Piatkowski/etal/2013a}, and computer vision \cite{Yin/Collins/2007a}. Thus, discrete graphical models build the foundation for various classes of machine techniques. 

For structures with high-order interactions, probabilistic inference is particularly challenging in graphical models defined over discrete variables, for which computation of the normalizing constant (and thus the data likelihood) is in general computationally intractable. A common way to circumvent the explicit normalization of the probability mass function is to compute the quantities of interest based on samples which are drawn from the graphical model. 
The problem of generating samples from graphical models traces back to the seminal work on
the Metropolis-Hastings algorithm \cite{Metropolis/etal/1953a,Hastings/1970a} and Gibbs-sampling \cite{Geman/Geman/1984a}, and as of today Markov chain Monte Carlo (MCMC) methods are still at the center of attention when samples are be generated from high-dimensional graphical models. 
MCMC methods are iterative---an initial guess is randomly modified repeatedly until the chain converges to the desired distribution. The actual time until convergence (mixing) is model dependent, hard to derive, and exponential in the number of variables \cite{Bubley/Dyer/1997a}. A more recent, promising line of research relies on random perturbations on the model parameters. These perturb-and-MAP (PAM) techniques \cite{Hazan/etal/2013a} compute the maximum a posterior (MAP) state of a graphical model whose potential function is perturbed by independent samples from a Gumbel distribution. The resulting perturbed MAP state can be shown to be an unbiased independent sample from the underlying graphical model. Assigning the correct Gumbel noise and solving the MAP problem are both exponential in the number of variables. Efficient perturbations have been discovered \cite{Niepert/etal/2021a}, sacrificing the unbiasedness of samples while delivering viable practical results. However, the exponential time complexity of MAP computation renders the method intractable in the worst-case. 

In this work, we propose a method for generating unbiased samples from graphical models on a quantum processors. Instead of an interative construction of samples as in MCMC or a perturbation as in PAM, our method generates samples from the probabilistic nature of a collapsing quantum state. However, since there is no free lunch, the success probability of our method decreases exponentially with the number of maximal cliques of the underlying conditional independence structure. As opposed to MCMC or PAM samplers, our method is a Las-Vegas type randomized algorithm. That is, it has an additional output that indicates if a generated samples was created successfully or if it has to be discarded.

Indeed, quantum algorithms for learning and inference of specific probabilistic models have been proposed, including quantum Bayesian networks \cite{Low/etal/2014a}, quantum Boltzmann machines \cite{QBMAmin18, QBMWiebe17, Wiebe2019GenerativeTO, Zoufal/etal/2020a}, and Markov random fields \cite{Zhao/etal/2021a,Bhattacharyya/2021a,Nelson/etal/2021a}. However, many of these methods are either approximate or require so-called {\em fault-tolerant} quantum computers---a concept that cannot yet be realized with the state-of-the-art quantum hardware. Instead, we derive the first quantum circuit construction that is exact and compatible with current quantum computing hardware. To prove the practical viability of our approach, we provide experimental results on a quantum simulation as well as actual quantum hardware, showing that our method can reliably carry out sampling and parameter learning on quantum computers.

\section{Problem formulation}
In this section, we formalize the problem of generating samples from a graphical model with a quantum circuit. %

\subsection{Parametrized family of models}
We consider positive joint probability distributions over $n$ discrete variables $\bX_v$ with realizations $\bx_v\in\cX_v$ for $v\in\{1,\dots,n\}$. The set of variable indices $V$ is referred to as {\em vertex set} and its elements $v\in V$ as {\em vertices}. Without loss of generality, the positive probability mass function (pmf) of the $n$-dimensional random vector $\bX$ can be expressed as\begin{equation}\label{eq:expfam}
\mP_{\bt,\phi}(\bX=\bx) = \frac{1}{Z(\bt)} \exp\left(\sum_{j=1}^d\bt_j \phi_j(\bx) \right)
\end{equation}
where $\phi=(\phi_1,\dots,\phi_d)$ is a set of {\em basis functions} or {\em sufficient statistics} that specify a family of distributions and $\bt\in\mR^d$ are parameters that specify a model within this family. When $\phi$ is clear from the context, we simply write $\mP_{\bt}$ and drop the explicit dependence on $\phi$.
The quantity $Z(\bt)$ denotes the model's partition function and is required for normalization such that $\mP$ becomes a proper probabiliy mass function. 
When $\Phi$ covers the conditional independence structure of $\bX$, \eqref{eq:expfam} can be rewritten as
\begin{equation}\label{eq:mrf}
\mP_{\bt}(\bX=\bx) = \frac{1}{Z(\bt)} \prod_{C\in\cC} 
\exp\left(\sum_{\by\in\cX_C} \bt_{C,\by} \phi_{C,\by}(\bx) \right)
= \frac{1}{Z(\bt)} \prod_{C\in\cC} \psi_C(\bx_C)
\end{equation}
where $\cC$ denotes the set of maximal cliques, i.e., a sub-set of $V$, for some arbitrary but fixed undirected graphical structure $G$ over $V$. 
The equality between \eqref{eq:expfam} and \eqref{eq:mrf} is known as Hammersley-Clifford theorem \cite{Hammersley/Clifford/1971a}. Setting\begin{equation}\label{eq:overstat}
    \phi_{C,\by}(\bx) = \prod_{v\in C} \mathbbm{1}(\bx_v = \by_v)
\end{equation}
is sufficient for representing any arbitrary pmf with conditional independence structure $G$ \cite{Pitman/1936a,Besag/1975a,Wainwright/Jordan/2008a}. In this case, the graphical model is called {\em Markov random field}. Moreover, $\phi(\bx)=(\phi_{C,\by}(\bx):C\in\cC, \by\in\cX_C)$ represents an {\em overcomplete} family, since there exists an entire affine subset of parameter
vectors $\bt$, each associated with the same distribution. Non-overcomplete, e.g., minimal, statistics facilitate uniqueness of models and are hence in favor when analyzing the learning process. Overcomplete families appear frequently when designing and analyzing algorithms for probabilistic inference \cite{Wainwright/Jordan/2008a}. 

\subsection{Sampling and the quantum state preparation problem}
\label{sec:sampling_and_quantum_state}
Generating samples from \eqref{eq:expfam} via plain inverse transform sampling is intractable as there are $\Omega(2^n)$ distinct probabilities involved. 
However, this exponentially large set has a native representation in terms of {\em quantum states}.

In quantum computing, information is represented via a quantum state $\ket{r}$ living in a {\em qubit register} of some fixed width $w$. While an ordinary (classical) $w$-bit register $\br$ stores a single bit string of length $w$, the corresponding quantum state $\ket{r}$ stores a full joint probability distribution over all possible $2^w$ bit strings, referred to as {\em basis states}. Quantum processors do not grant access to the raw probabilities. Instead, the qubit register can be {\em measured} to yield one specific bit string $\bx\in\{0,1\}^w$--corresponding to one of the $2^w$ basis states $\ket{\boldsymbol{x}}$. Notably, each possible bit string is measured with the probability that is stored in the quantum state. 
Transferring some initial qubit state $\ket{r_{\operatorname{in}}}$ to some desired output state $\ket{r_{\operatorname{out}}}=\bC \ket{r_{\operatorname{in}}}$ can be achieved, e.g., by application of a unitary quantum circuit $\bC$ acting on all $w$ qubits. 
Any unitary operator $U$ satisfies $U^\dagger U=U U^\dagger=I$ and its eigenvalues have modulus $1$. Here, $I$ denotes the identity and $\dagger$ denotes the conjugate transpose.
Quantum states are always $\ell_2$ normalized. This normalization is preserved by unitary operations. 
The probability for measuring a specific bit string $\bx$ as the output of the circuit is given by the Born rule: \begin{equation}\label{eq:born}
\mP_{\bC}(\bx) = |\braket{\bx | r_{\operatorname{out}}}|^2 = |\bra{\bx} \bC \ket{r_{\operatorname{in}}}|^2,  
\end{equation}
where $\bra{\bx}=\left(\ket{\bx}\right)^{\dagger}$.

In practice, one often uses one or two-qubit unitaries. These operations can be composed via matrix multiplication and Kronecker products to form higher-order qubit transformations. %
Borrowing terminology from digital computing, unitary operators acting on qubits are also called \emph{quantum gates}.
In the context of this work, we will be especially interested in the gates \[
  X = \begin{bmatrix}
0 & 1\\
1 & 0
\end{bmatrix}
,\hspace{0.25cm}
  Z = \begin{bmatrix}
1 & 0\\
0 & -1
\end{bmatrix}
,\hspace{0.25cm}
  I = \begin{bmatrix}
1 & 0\\
0 & 1
\end{bmatrix},
\] and $H=\frac{1}{\sqrt{2}}(X+Z)$. The matrices $X$, $Z$, and $I$ are so-called \textit{Pauli matrices} and $H$ represents the \textit{Hadamard} gate, with $\ket{+} := H\ket{0} = \textstyle{\frac{1}{\sqrt{2}}}\left( \ket{0} + \ket{1}\right)$. 
Higher-order operators are derived by taking the Kronecker product of the above, e.g., $X\otimes I^{\otimes 2}$. %
Finally, the action of any quantum circuit $\bC$ can be written as a product of unitaries: $\bC = U_d U_{d-1} U_{d-2} \dots U_1$, 
where $d$ is the {\em depth} of the circuit.
It is important to understand that a gate-based quantum computer receives its circuit \emph{symbolically}, typically as a sequence of Kronecker and matrix products of low dimensional unitaries---the implied $2^n \times 2^n$ matrix \emph{does not have to be materialized}.
A detailed introduction into this topic can be found in \cite{Nielsen/Chuang/2016a}. 

\begin{definition}[Graphical model quantum state preparation problem]
Given any discrete graphical model over $n$ binary variables, defined via $(\bt,\phi)$, find a quantum circuit $\bC$ which maps an initial quantum state $\ket{r_{\operatorname{in}}}$ to an output state $\ket{r_{\operatorname{out}}}$ such that\begin{equation}\label{eq:prob}
    \mP_{\bC} = 
\mP_{\bt}
\end{equation}
as specified by \eqref{eq:expfam} and \eqref{eq:born}. 
\end{definition}

In what follows, we explain how to find $\bC$ that solves \eqref{eq:prob} for an appropriate $\ket{r_{\operatorname{in}}}$.

\section{Main Results}
We devise a quantum algorithm in which each vertex of a graphical model over binary variables is mapped to one qubit of a quantum circuit. In addition, $1+|\cC|$ {\em auxiliary qubits} are required to realize specific operations as explained below. 
Our result consists of two parts.
First, we present a derivation of the Hamiltonian $H_{\bt}$ encoding the un-normalized, negative log-probabilities. Then, we employ $H_{\bt}$ to construct a quantum circuit that allows us to draw unbiased and independent samples from the respective graphical model.

\subsection{The Hamiltonian}
We start by transferring the sufficient statistics of the graphical model family into a novel matrix form. This allows us to construct a Hamiltonian matrix that encodes the parameters $\bt$ as well as the conditional independence structure $G$. 

\begin{definition}[Pauli-Markov Sufficient Statistics]\label{def:pm}
Let $\phi_{C,\by}:\cX\to\mR$ for $C\in\cC$ and $\by\in\cX_C$ denote the sufficient statistics of some overcomplete family of graphical models. %
The diagonal matrix $\Phi_{C,\by} \in\{0,1\}^{2^n \times 2^n}$, defined via
$(\Phi_{C,\by})_{i,j} = \phi_{C,\by}(\bx^j)$ iff $i=j$ (and $0$ otherwise), 
denotes the Pauli-Markov sufficient statistics. Where $\bx^j$ denotes the $j$-th full $n$-bit joint configuration w.r.t.~some arbitrary but fixed order.
\end{definition}

A naive computation of the Pauli-Markov sufficient statistics for any fixed $(C,\by)$-pair is intractable due to the sheer dimension of $\Phi_{C,\by}$. However, it turns out that the computation can be carried out with a linear number of Kronecker product evaluations via Alg.~\ref{alg:paulimarkov}. Thus, a \emph{symbolic} representation of $\Phi_{C,\by}$ can be computed in linear time.
\begin{theorem}[Computing Sufficient Statistics]\label{thm:pmcorrectness}
Let $\Phi_{C,\by}^v$ be the intermediate result of Algorithm~\ref{alg:paulimarkov} after iteration $v$. Then, $(\Phi^v_{C,\by})_{j,j} = \phi_{C\cap[v],\by}(\bx^j_{[v]})$. %
\end{theorem}
The reader will find the proof of Theorem~\ref{thm:pmcorrectness} in Section \ref{proof:pmcorrectness} of the Supplementary Material. 

\begin{algorithm}[t!]
\caption{Computing Pauli-Markov sufficient statistics in linear time.}
\label{alg:paulimarkov}
\begin{algorithmic}[1]
\REQUIRE $C\subseteq V$, $\by \in \cX_C$
\ENSURE $\Phi_{C,\by} = \Phi$
\STATE $\Phi \leftarrow 1$
\FOR{$v \in V$}
\IF{$v\not\in C$}
\STATE $\Phi$ $\leftarrow$ $\Phi \otimes I$ 
\ELSIF{$v \in C$ and $\by_v = 1$}
\STATE $\Phi$ $\leftarrow$ $\Phi \otimes (I-Z)/2$ 
\ELSIF{$v \in C$ and $\by_v = 0$}
\STATE $\Phi$ $\leftarrow$ $\Phi \otimes (I+Z)/2$ 
\ENDIF
\ENDFOR
\RETURN $\Phi$
\end{algorithmic}
\end{algorithm}

Obviously, the symbolic representation of the tensor product computed by Alg.~\ref{alg:paulimarkov} has length $\Theta(n)$. Hence, the algorithm runs in linear time. 
Given the notion of Pauli-Markov statistics, the first part of our main result is stated in the following theorem.

\begin{theorem}[Hamiltonian]\label{thm:hamiltonian}
Assume an overcomplete binary graphical model specified by $(\bt,\phi)$. When $
H_{\bt} = -\sum_{C\in\cC} \sum_{\by\in\cX_C} \bt_{C,\by} \Phi_{C,\by}$, 
then $\mP_{\bt}(\bx^j)=({\exp_M(-H_{\bt})}/{\otrace{\exp_M(-H_{\bt})}})_{j,j}$
where $\exp_M$ is the matrix exponential and $\otrace$ the trace. 
\end{theorem}
The reader will find the proof of Theorem~\ref{thm:hamiltonian} in Section \ref{proof:hamiltonian} of the Supplementary Material.

Thus, $H_{\bt}$ accumulates the conditional indepence structure $G$ of the underlying random variable via $\Phi_{C,\by}$ as well as the model parameters $\bt$. Clearly, $H_{\bt}$ is not unitary, since it is real and hence its eigenvalues cannot have modulus $1$ when $H_{\bt}\not=I^{\otimes n}$. Thus, an unitary embedding is required to realize $H_{\bt}$ as part of a quantum circuit. %

\subsection{The Circuit}
\label{sec:circuit}
Using the Hamiltonian $H_{\bt}$ from the previous section, we now construct a circuit that implements the non-unitary operation $\exp{\left(- H_{\bt}\right)}$, based on unitary embeddings, a special pointwise polynomial approximation, and the factorization over cliques. We use this quantum circuit to construct a quantum state whose sampling distribution  %
is proportional to that of any desired graphical model over binary variables. 
Our findings are summarized in the following theorem.

\begin{theorem}[Quantum Circuit for Discrete Graphical Models]\label{thm:circ}
Given any overcomplete discrete graphical model over $n$ binary variables, defined via $(\bt,\phi)$. There exists an $\cO(d)$-depth quantum circuit $\bC_{\bt}$ over $m=n+1+|\cC|$ qubits that prepares a quantum state whose sampling distribution is equivalent to the graphical model such that
$\mP_{\bt}(\bx) \propto \mP_{\bC}(\bx)  $
for each $\bx\in\{0,1\}^n$ with $ \mP_{\bC}(\bx)  = \sum_{z\in\set{0,1}}|(\bra{0}^{\otimes |\cC|}\otimes\bra{z}\otimes\bra{\bx})\bC_{\bt}
\ket{+}^{\otimes m}|^2
$ .
\end{theorem}
The reader will find the proof of Theorem~\ref{thm:circ} in Section \ref{proof:circ} of the Supplementary Material. 

\begin{figure}[t!]
\hspace*{-0.5cm}
\scalebox{0.8}{
\Qcircuit @C=1.0em @R=0.2em @!R { \\
	 	\nghost{{v}_{0} :  } & \lstick{{v}_{0} :  } & \gate{\mathrm{H}} \barrier[0em]{5} & \qw & \qw & \multigate{3}{\mathrm{U^A}} & \multigate{3}{\left(\mathrm{U^A}\right)^\dagger} & \qw & \qw \barrier[0em]{5} & \qw & \qw & \multigate{3}{\mathrm{U^B}} & \multigate{3}{\left(\mathrm{U^B}\right)^\dagger} & \qw & \qw \barrier[0em]{5} & \qw & \meter & \qw & \qw & \qw & \qw\\
	 	\nghost{{v}_{1} :  } & \lstick{{v}_{1} :  } & \gate{\mathrm{H}} & \qw & \qw & \ghost{\mathrm{U^A}}& \ghost{\left(\mathrm{U^A}\right)^\dagger} & \qw & \qw & \qw & \qw & \ghost{\mathrm{U^B}} & \ghost{\left(\mathrm{U^B}\right)^\dagger} & \qw & \qw & \qw & \qw & \meter & \qw & \qw & \qw\\
	 	\nghost{{v}_{2} :  } & \lstick{{v}_{2} :  } & \gate{\mathrm{H}} & \qw & \qw & \ghost{\mathrm{U^A}} & \ghost{\left(\mathrm{U^A}\right)^\dagger} & \qw & \qw & \qw & \qw & \ghost{\mathrm{U^B}} & \ghost{\left(\mathrm{U^B}\right)^\dagger} & \qw & \qw & \qw & \qw & \qw & \meter & \qw & \qw\\
	 	\nghost{{a}_{3} :  } & \lstick{{a}_{3} :  } & \gate{\mathrm{H}} & \qw & \qw & \ghost{\mathrm{U^A}} & \ghost{\left(\mathrm{U^A}\right)^\dagger} & \qw & \qw & \qw & \qw & \ghost{\mathrm{U^B}} & \ghost{\left(\mathrm{U^B}\right)^\dagger} &\qw & \qw & \qw & \qw & \qw & \qw & \qw & \qw\\
	 	\nghost{{a}_{4} :  } & \lstick{{a}_{4} :  } & \qw & \qw & \gate{\mathrm{H}} & \ctrl{-1} & \ctrlo{-1} & \gate{\mathrm{H}} & \meter & \qw & \qw & \qw & \qw & \qw & \qw & \qw & \qw & \qw & \qw & \qw & \qw\\
	 	\nghost{{a}_{5} :  } & \lstick{{a}_{5} :  } & \qw & \qw & \qw & \qw  & \qw & \qw & \qw & \qw & \gate{\mathrm{H}} & \ctrl{-2} & \ctrlo{-2} & \gate{\mathrm{H}} & \meter & \qw & \qw & \qw & \qw & \qw & \qw\\
	 	\nghost{\mathrm{c :  }} & \lstick{\mathrm{c :  }} & \lstick{/_{_{5}}} \cw & \cw &  \cw & \cw & \cw & \cw & \dstick{_{_{\hspace{0.0em}4}}} \cw \ar @{<=} [-2,0] & \cw &  \cw & \cw & \cw & \cw & \dstick{_{_{\hspace{0.0em}5}}} \cw \ar @{<=} [-1,0] & \cw & \dstick{_{_{\hspace{0.0em}0}}} \cw \ar @{<=} [-6,0] & \dstick{_{_{\hspace{0.0em}1}}} \cw \ar @{<=} [-5,0] & \dstick{_{_{\hspace{0.0em}2}}} \cw \ar @{<=} [-4,0] & \cw & \cw\\
\\ }}
\caption{Exemplary quantum circuit $\bC_{\bt}$ as specified in Eq.~\ref{eq:qcfinal} with $U^C = \prod_{\by\in\cX_C} U^{C,\by}({\bt_{C,\by}})$. In this example, the underlying graph has vertex set $V=\{v_0,v_1,v_2\}$ and clique set $\cC=\{A,B\}$. The circuit requires $|\cC|+1=3$ auxiliary qubits, one for the unitary embedding $U_j$ of the sufficient statistic and one for the real part extraction of each clique. The auxiliary qubits $a_4$ and $a_5$ are measured before the circuit has been fully evaluated. This allows for an early restart whenever real part extraction fails.\label{fig:qcmrf}}
\end{figure}
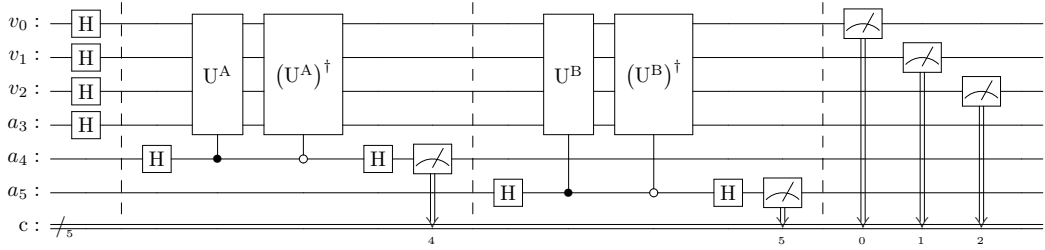
The construction is based on unitary embeddings $U^{C,\by}({\bt_{C,\by}})$ of $\exp(\bt_{C,\by} \Phi_{C,\by})$. 
An exemplary circuit is shown in Fig.~\ref{fig:qcmrf}.
The first $n$ qubits of the circuit realize the target register $\ket{\bx}$, that represents the $n$ binary variables of the graphical model. The latter $1+|\cC|$ qubits represent an auxiliary register $\ket{\ba}$, which is required for the unitary embedding and the extraction of real parts as described in Section \ref{proof:pmcorrectness} of the Supplementary Material. 
The Hadamard gates at the beginning are required to bring all qubits into the state $\ket{+}^{\otimes m}$, as described in Sec.~\ref{sec:sampling_and_quantum_state}. This state realizes a uniform sampling distribution over $\{0,1\}^m$ in which any measurement result has the same probability. The unitaries $U^{C}$ are then manipulating this state such that it becomes proportional to $\mP_{\bt}$ on the target register $\ket{\bx}$.

Note the proportionality (as opposed to equality) between $\mP_{\bt}$ and the output state of the circuit $\bC_{\bt}$. The measurements from the quantum circuit are taken from $\ket{\ba}\otimes\ket{\bx}$, i.e., from the joint distribution over auxiliary and target qubits. Samples from the graphical model can, then, be extracted from the quantum measurements by conditioning on the event that the last $|\cC|$ qubits are all measured as $0$.
In practice, one may discard all quantum measurements with non-zero auxiliary qubits to generate unbiased samples from the underlying graphical model.

Notable, our circuit construction shares a defining property of undirected graphical models.
\begin{corollary}[Unitary Hammersley-Clifford]
Setting $U^C(\bt_C) = (H_C \otimes I^{\otimes (n+1)}) \prod_{\by\in\cX_C} U^{C,\by}({\bt_{C,\by}})$ reveals the clique factorization $\bC_{\bt} = \prod_{C\in\cC}  U^C(\bt_C) $ as predicted by the Hammersley-Clifford theorem \cite{Hammersley/Clifford/1971a}.
\end{corollary}
We like to stress that the clique factorization is of utmost importance for our derivation. Without exploiting the factorization, computation of $\exp(- H)$ must be carried out directly. In that case, the pointwise polynomial approximation of the exponential function that we utilize in the proof of Theorem~\ref{thm:circ} would no longer suffice and we would have to resort to a uniform polynomial approximation of $\exp$ over some interval---introducing a polynomial approximation error and increasing the depth of the overall circuit proportional to the polynomial degree.

\section{Inference}\label{sec:inference}
The four main inference tasks that can be done with graphical models are (i) sampling, (ii) MAP inference, (iii) parameter learning, and (iv) estimating the partition function. The ability to generate samples from the graphical model follows directly from Theorem~\ref{thm:circ}. Here, we provide the foundations required to address inference tasks (ii)-(iv) based on our circuit construction. 
\subsection{MAP Estimation}
Computing the MAP state of a discrete graphical model is required when the model serves as the underlying engine of some supervised classification procedure. More precisely, the MAP problem is $
\bx^* = \operatorname{arg\,max}_{\bx \in \cX} \mP_{\bt}(\bx) = \operatorname{arg\,max}_{\bx \in \cX} \bt^\top \phi(\bx)
$. 
Theorem~\ref{thm:hamiltonian} asserts that our Hamiltonian $H_{\bt}$ carries $-\bt^\top \phi(\bx)$ for all $\bx\in\cX$ on its diagonal. Since $H_{\bt}$ is itself diagonal, $-\bt^\top \phi(\bx^*)$ is the smallest eigenvalue of $H_{\bt}$. Computing the smallest eigenvalue and the corresponding eigenvector---which corresponds to $\bx^*$ in the $2^n$-dimensional state space---is a well studied \textbf{QMA}-hard problem in the quantum computing community. Heuristic algorithms like the variational quantum eigensolver \cite{PeruzzoVQE2014}
can, thus, be directly applied to our Hamiltonian in order to approximate the MAP state.

\subsection{Parameter Learning}\label{sec:learn}
Parameters of the graphical model can be learned consistently via the maximum likelihood principle. Given some data set $\cD$ that contains samples from some desired distribution $\mP^*$, we have to minimize the convex objective $\ell(\bt)=-(1/|\cD|)\sum_{\bx\in\cD} \log \mP_{\bt}(\bx)$ with respect to $\bt$. Differentiation reveals $\nabla\ell(\bt)=\hat{\bmu}-\tilde{\bmu}$ where $\tilde{\bmu}=(1/|\cD|)\sum_{\bx\in\cD}\phi(\bx)$ and $\hat{\bmu}=\sum_{\bx\in\cX}\mP_{\bt}(\bx) \phi(\bx)$. The latter quantity can be estimated by drawing samples from the graphical model. Notably, our circuit does not require a quantum-specififc training procedure, since the circuit $\bC_{\bt}$ is itself parametrized by the canonical parameters $\bt$ of the corresponding family. This allows us to run any iterative numerical optimization procedure on a digital computer and employ our circuit as sampling engine for estimating $\hat{\bmu}$. After each training iteration, we update the parameters of the circuit and are ready to generate the samples for the next iteration. 
We have to remark that our circuit allows for an alternative way to estimate the parameters. The circuit can be parametrized by a vector of rotation angles $\bgam$. Thus, we may learn $\bgam$ instead, utilizing a quantum gradient $\nabla_{\bgam}^{\bx}$ where
\begin{equation*}
\left(\nabla_{\bgam}^{\bx}\right)_j = \textstyle{\frac{\partial}{\partial \bgam_j}} \sum\limits_{i\in\set{0,1}}|(\bra{0}^{\otimes |\cC|}\otimes\bra{i}\otimes\bra{\bx})\bC_{\bgam}
\ket{+}^{\otimes m}|^2
\end{equation*}
\cite{qGANDallaire_Demers18, zoufal2021generativeQML}. The corresponding canonical parameters can be recovered from $\bgam$ via $\bt_j = 2\log\cos(2\bgam_j)$. 

\subsection{Approximating the Partition Function}
Estimating the partition function $Z(\bt)$ of a graphical model allows us to compute the probability of any desired state directly via the exponential family form $\mP_{\bt}(\bx)=\exp(\bt^\top\phi(\bx)-\ln Z(\bt))$.
Computing $Z(\bt)$ is a well recognized problem, not least because of its sheer complexity---the problem is \textbf{\#P}-hard. 
It turns out that we can get $\tau(\theta)=2^{-n}2\exp(\ln Z(\bt))$ with probability at least $3/4$ and $\cO(\log \nicefrac{1}{\varepsilon})$ extra auxiliary qubits by modifying our circuit construction to yield a $\cO(d(\operatorname{poly}(n)+1/(\varepsilon\sqrt{\tau(\bt)})))$-depth circuit. The basic idea is to apply quantum trace estimation as defined in \cite[Theorem 7]{Bravyi/etal/2021a} to the matrix from Eq.~\ref{eq:realfactorization} (Provided in Section \ref{proof:circ} of the Supplementary Material). %
Due to the high depth, the resource consumption of this procedure is prohibitive for current quantum computers. Nevertheless, it opens up avenues for probabilistic inference on upcoming fault-tolerant quantum hardware. 
\renewcommand{\arraystretch}{1.15}
\begin{table*}[t!]
\caption{Median fidelities ($F$) over 10 runs, computed from $N=100000$ samples from our method -- quantum circuits for graphical models (QCGM) -- on the QASM simulator as well as an IBM Falcon quantum processor. The same number of samples has been generated via Gibbs-sampling and PAM. For QCGM, we also report the number of effective samples $\tilde{\delta}^*N$, where $\tilde{\delta}^*$ is the empirical success probability. %
\label{tab:results12}}

\begin{tabular*}{\textwidth}{c @{\extracolsep{\fill}} llccccccc}
\toprule\\[-0.35cm]
&&\model{1.0}{0}&\model{1.0}{1}&\model{1.0}{2}&\model{1.0}{3}&\model{1.0}{6}&\model{1.0}{7}&\model{1.0}{8}\\
\midrule
QCGM
&$F$&1.000&1.000&0.998&0.874&1.000&0.999&0.999&\\
{\footnotesize Simulation}
&$\tilde{\delta}^*N$&1864&14924&5055&25&9961&9315&11485&\\
\midrule
QCGM
&$F$&0.998&0.903&0.790&0.596&0.875&0.588&0.742&\\
{\footnotesize Hardware}
&$\tilde{\delta}^*N$&17229&23614&9800&11889&35343&24298&52768&\\
\midrule
Gibbs
&$F$&1.000&0.999&0.994&0.987&0.994&0.950&0.982&\\
\midrule
PAM
&$F$&0.999&0.999&0.969&0.939&0.999&0.957&0.999&\\
\bottomrule
\end{tabular*}
\end{table*}
\setlength{\belowcaptionskip}{-0.5cm}
\begin{figure}[t!]
    \centering
    \includegraphics[width=0.45\textwidth]{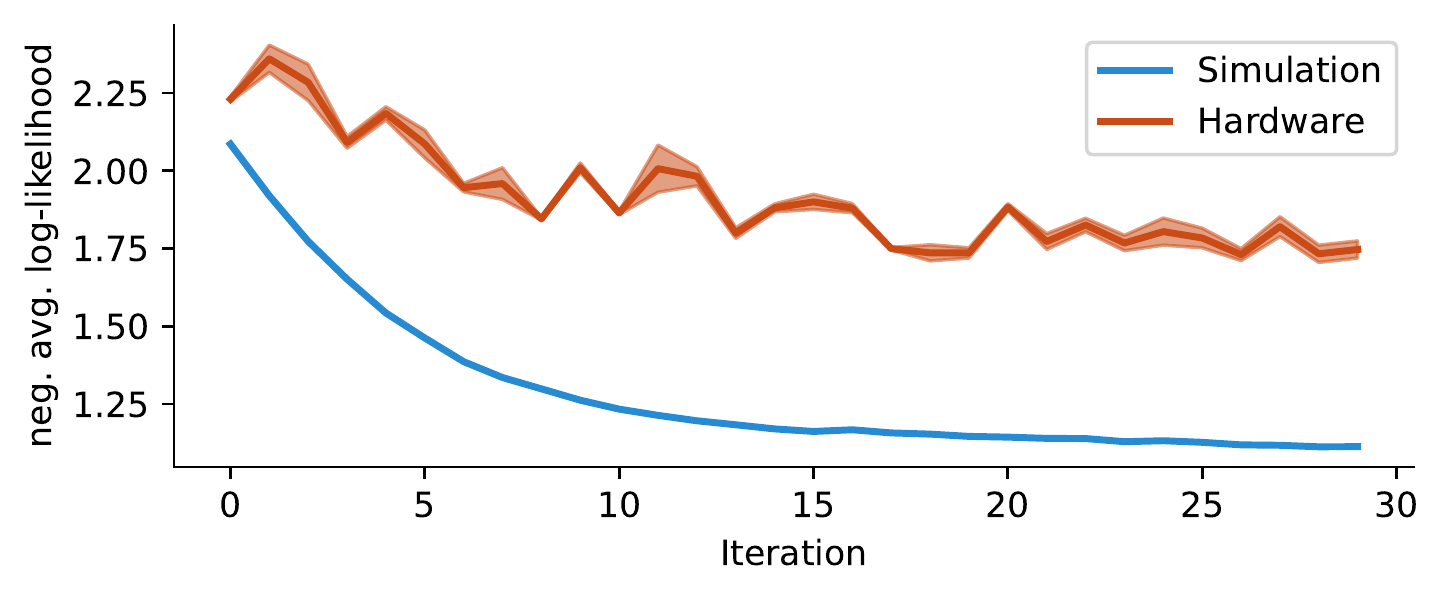}\hspace{0.5cm}
    \includegraphics[width=0.45\textwidth]{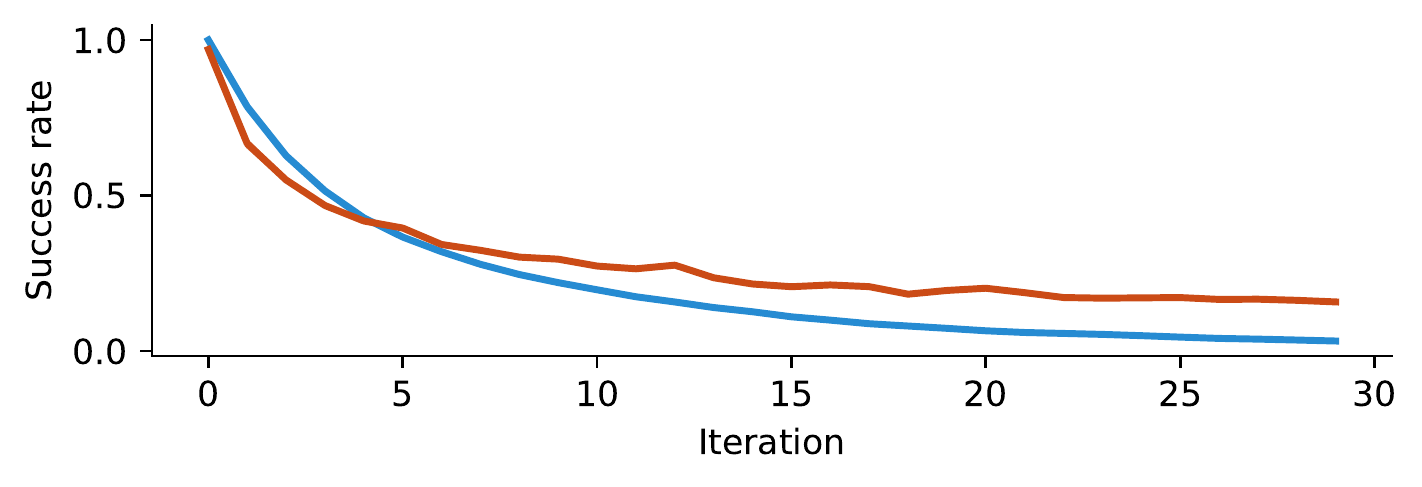}
    \caption{Left: Learning curve for QCGM over 30 ADAM iterations on the Qiskit QASM simulator and an actual quantum processor. %
    Right: The empirical success rate $\tilde{\delta}^*$ of each training iteration.%
    }
    \label{fig:results3}
\end{figure}

\section{Limitations}\label{sec:limits}
The real part extraction, utilized in Theorem~\ref{thm:circ} might fail with some probability $1-\delta_U$ that depends on the specific unitary $U$. The practical impact of this fact is studied in the experiments. However, measuring the extra qubits tells us if the real part extraction succeeded. We may hence repeat the procedure until we observe a success. The overall success probability of the circuit $\bC_{\bt}$ is denoted as  $\delta^*=\prod_{C\in\cC} \delta_C$, where $\delta_C$ is the success probability for extracting the real part of clique $C$. This quantity is a-priori unknown and depends on the specific graphical model. More specifically, it decreases exponentially in the number of maximal cliques.
The general class of quantum circuits having this property is also called repeat-until-success circuit.
In fact, computing the actual success probability requires a full and hence exponentially expensive simulation of the quantum system. As an alternative, we may simply run the circuit for a number of repetitions to obtain an empirical estimate $\tilde{\delta}$ to $\delta^*$, as we do in the experimental evaluation. 

Assuming that $\delta^*$ (or its estimate $\tilde{\delta}$) is not sufficiently large, amplification techniques \cite{Brassard2000QuantumAA, %
Grover/2005a,Yoder/etal/2014a, Gilyen/etal/2019a} can help us to increase the success probability at the cost of additional auxiliary qubits or a higher circuit depth. %
Specifically, applying singular value transformation \cite{Gilyen/etal/2019a} to the real part extraction can raise $\delta^*$ to $1 - \varepsilon$ for any desired $\varepsilon>0$ with additional depth $\Omega(\log(1/\varepsilon)\sqrt{\delta})$ per clique. 

Moreover, the number of required auxiliary qubits---one per clique---can be prohibitive for the realization of large models with current quantum computers. 
It should hence be noted that the number of auxiliary qubits may be reduced with intermediate measurements. Since the real part extractions %
are applied in series, one may use a single auxiliary qubit for all real part extractions which is measured and reset to $\ket{0}$ after every single extraction. %
Intermediate measurements are, however, still technically challenging on current quantum devices and can, in fact, increase the system noise.
 
 Due to limited coherence times and physical noise induced by cross-talk between qubits, increasing the qubit number or the circuit depth makes it harder to run the algorithm with near-term quantum hardware. We refer to Section \ref{sec:qcapx} in the Supplementary Material for an explanation of terms that are specific to quantum computing. 
Thus, we do neither apply intermediate measurements nor probability amplification in order to ensure feasibility of our approach on actual quantum computing hardware. 

The hardware related limitations of our method can be summarized as follows.
\begin{theorem}[Resource Limitations]
The circuit construction from Theorem \ref{thm:circ} requires $|\cC|+1$ extra qubits. The expected runtime until a valid sample is generated is $\cO(1/\delta_{\min}^{|\cC|})$ with $\delta_{\min}=\min_{C\in\cC} \delta_C$, and hence, exponential in the number of cliques.
\end{theorem}

Finally, our method assumes a discrete graphical model family with binary variables and overcomplete sufficient statistics. Nevertheless, any discrete family with vertex alphabets of size $k$ can be transformed into an equivalent family with $\cO(n \log_2 k)$ binary variables. Clearly, increasing the number of variables increases the number of required qubits, which complicates the execution of our method on actual quantum processors. 

Moreover, any model family with minimal statistic can be converted into one with overcomplete statistics \cite{Wainwright/Jordan/2008a}. We hence treat this limitation as not crucial.

\section{Experimental Evaluation}
\label{sec:exp_eval}
Here, we want to evaluate the practical behavior of our method, named QCGM, by answering a set of questions which are discussed below.
Theorem~\ref{thm:circ} provides the guarantee that the sampling distribution of our QCGM is identical to $\mP_{\bt}$ of some given discrete graphical model. However, actual quantum computers are not perfect and the computation is influenced by various sources of quantum noise, each having an unknown distribution \cite{Nielsen/Chuang/2016a}. %
Hence, we investigate: 
\textbf{(Q1)} How close is the sampling distribution of QCGM on actual state-of-the-art quantum computing devices to a noise-free quantum simulation, Gibbs-sampling, and perturb-and-MAP? 
According to Sec.~\ref{sec:real}, measuring the auxiliary qubits of $\bC_{\bt}$ tells us if the real part extraction has failed or not.
The actual success probability is however unknown. The second question we address with our experiments is hence: %
\textbf{(Q2)} What success probability should we expect and what parts of the model influence $\tilde{\delta}^*$?
Third, as explained in Section~\ref{sec:learn}, the parameter learning of QCGM can be done analogously to the classical graphical model, based on a data set $\cD$ and samples $\cS$ from the circuit. As explained above, samples from the actual quantum processor will be noisy. However, it is known since long that error-prone gradient estimates can still lead to reasonable models as long as inference is carried out via the same error-prone method \cite{Wainwright/2006a}. Our last question is thus: %
\textbf{(Q3)} Can we estimate the parameters of a discrete graphical model in the presence of quantum noise? %

\subsection{Experimental Setup}\label{sec:exsetup}
For question $\textbf{(Q1)}$, we design the following experiment. First, we fix the conditional independence structures shown in Tab.~\ref{tab:results12}. For each structure, we generate 10 graphical models with random parameter vectors drawn from $\mU[-5,0)^d$ which allows for a rather large dynamic range of the random model. 
QCGM is implemented using Qiskit \cite{qiskit} (available under Apache License 2.0) and realized by applying the circuit $\bC_{\bt}$ to the state $\ket{+}$. The probabilities for sampling $\bx$ are evaluated by taking $N$ samples from the prepared quantum state and computing the relative frequencies of the respective $\bx$.%
The quantum simulation is carried out by the Qiskit QASM simulator which is noise-free. Any error that occurs in the simulation runs is thus solely due to sampling noise, i.e., due to the fact that we draw a finite number of samples. The experiments on actual quantum computers are carried out on five devices, each being a 27-qubit IBM Falcon superconducting quantum processor \cite{ibmQX}, and employ tensored error mitigation \cite{2021TensErrorMitigation}. Gibbs-sampling is performed with a fixed burn-in of $b=100$ samples. Moreover, each variable is re-sampled $n$ times and we discard $b$ samples between each two accepted Gibbs-samples to enforce independence. However, these choices are heuristics and prone to error. Lastly, we apply perturb-and-MAP sampling  \cite{Papandreou/Yuille/2011a,Hazan/etal/2013a} with sum-of-gamma (SoG) perturbations \cite{Niepert/etal/2021a}. The SoG approach is superior to other inexact PAM approaches. However, since different clique factors will be perturbed with non-independent noise terms, each PAM sample comes from a biased distribution. 
$N=100000$ samples are drawn from each sampler, including the number of samples discarded by QCGM and Gibbs-sampling. 
The quality of each sampling procedure is assessed by the %
\emph{fidelity}, defined for two probability mass functions $\mP$ and $\mQ$ via $F(\mP,\mQ) = (\sum_{\bx\in\cX} \sqrt{\mP(\bx)\mQ(\bx)})^2$. 
$F$ is a common measure to assess the reliability of hardware qubits and quantum gates. Moreover, when $\cH(\mP,\mQ)$ is the Hellinger distance, then $\cH(\mP,\mQ)^2=1-\sqrt{F(\mP,\mQ)}$. %

For question $\textbf{(Q2)}$, we consider the very same setup as above, but instead of $F$, we compute  the empirical success rate of the QCGM as $\tilde{\delta}^*=\text{number of succeeded samplings}/N$. This is computed for each of the 10 runs on each quantum computer and the quantum simulator. 
Finally, for question $\textbf{(Q3)}$, we draw $N$ samples from a graphical model with edge set $\{(0,1),(1,2)\}$ via Gibbs-sampling. These samples are then used to train the parameters $\bt$ of a QCGM via ADAM \cite{Kingma/Ba/2015a} to compensate for the noisy gradient information.
For each of the 30 training iterations, we report the estimated negative, average log-likelihood and the empirical success rate. Training is initialized at $\bt=\boldsymbol{0}$. %

\subsection{Experimental Results}
The results of our experimental evaluation are shown in Tab.~\ref{tab:results12} and Fig.~\ref{fig:results3}. Numbers reported are median values over all runs. Regarding question \textbf{(Q1)}, we see that the fidelity of the simulation attains the maximal possible value---as predict by Theorem~\ref{thm:circ}---whenever enough samples could be gathered from the model. The limited number of effective samples $\tilde{\delta}^* N$ is due to the fact that we kept the number of sample trials $N$ constant over all experiments. Increasing $N$ for increased model complexity would mitigate this effect. Moreover, the fidelity on actual quantum hardware degrades as the model becomes more complex. It is important to understand that this is an artifact of the noise in current state-of-the-art quantum processors that will be reduced with improved quantum processor technology. When one considers the best result on each structure (and not the median), QCGM frequently attains fidelities of $>0.9$ also on quantum hardware. Finally, we see with quantum simulation that QCGM has the potential to outperform classical sampling methods with respect to the fidelity when enough samples are available. This can be explained by the fact that QCGM is guaranteed to return unbiased and independent samples from the underlying model. Gibbs-sampling can only achieve this if the hyper-parameters are selected carefully, or, in case of PAM, when an exact perturbation and an exact MAP solver are considered. For question \textbf{(Q2)}, we see from Tab.~\ref{tab:results12} that the success probability degrades when the number of maximal cliques increases. The sheer number of variables or model parameters does not have any impact, since perfect samples can be generated by the QCGM even for high-order models where the maximal clique size is $\geq 3$. Although the fidelities are better for the simulation, higher $\delta^*$ are observed with quantum hardware. We conjecture that this is due to hardware noise induced false sample acceptance. Interestingly, the second plot of Fig.~\ref{fig:results3} reveals that $\delta^*$ also degrades as a function of the model's entropy. Since we initialize the training with all elements of $\boldsymbol{\theta}$ being $0$, we start at maximum entropy. Parameters are refined during the learning and the entropy is reduced. However, the first plot of the figure shows that the training progresses for both, simulated and hardware results. The answer to $\textbf{(Q3)}$ is hence affirmative. 

\section{Conclusion}
We introduce an exact representation of discrete graphical models with a quantum circuit construction that acts on $n+1+|\cC|$ qubits and is compatible with current quantum hardware. This method enables unbiased, hyper-parameter free sampling while keeping the theoretical properties of the undirected model intact, e.g., our quantum circuit factorizes over the set of maximal cliques as predicted by the Hammersley-Clifford theorem. 
Although, our results are stated for binary models, equivalent results for arbitrary discrete state spaces can be derived, where multiple qubits encode one logical non-binary random variable. 
The full compatibility between the classical graphical model and our unitary embedding is significant, since it allows us to benefit from existing theory as well as quantum sampling.
A distinctive property of QCGM is that the algorithm itself indicates whether a sample should be discarded. 
Unlike related approaches, our method is based on a structured quantum circuit construction and is, thus, conjectured to be more robust against exponentially vanishing gradients than many QML approaches \cite{McClean_2018BarrenPlateaus}. 
The experiments conducted with numerical simulations and actual quantum hardware show that QCGMs perform well for certain conditional independence structures but suffer from small success probabilities for structures with large $|\cC|$.
In particular, in the latter case QCGM could significantly benefit from amplitude amplification techniques to boost the success probability.
 It remains open for future research to potentially remove the dependence of $\delta^*$ on the number of maximal cliques and to study the relationship of limiting factors between classical and quantum sampling methods. 
 In any case, our results open up new avenues for probabilistic machine learning on quantum computers by show-casing that the natural stochasticity of quantum models can be beneficial for a large model class. 

\section*{Social Impact}\label{sec:social}
We believe that this work, as presented here, does not exhibit any foreseeable societal consequences.

\newpage

\appendix

\section{State-of-the-Art in Quantum Computing}\label{sec:qcapx}

In recent years, the first generations of quantum computers became broadly available. The systems that are available today are so-called noisy quantum computers \cite{QV64Jurcevic20} which are still strongly impaired by physical limitations.
More explicitly, there are various sources of physical noise which impact a quantum circuit.

One problem is \emph{decoherence}. After a certain time -- which is currently still relatively short -- a qubit state cannot be maintained due to interaction with the environment. Hence, the information encoded in the qubit gets lost.  
Furthermore, various qubits in one quantum processor can influence each other in an undesired fashion. This is termed \emph{cross-talk}. 
The application of quantum operations (gates) and the readout of measurements are also impacted by physical disturbances and, thus, introduce noise in a quantum circuit.
If a circuit is, thus, to be executed on today's quantum hardware, the number of qubits and the number of consecutive quantum gates have to be limited. Otherwise, the impact of physical noise can be too strong to receive meaningful circuit results. 
In order to cope with those noise sources, researchers are investigating error mitigation \cite{2021TensErrorMitigation} and error correction \cite{FaultTolerantQCDevitt15} methods.
While the broad realization of error correction is still an open task for future research, error mitigation is already helpful in improving results from existing quantum hardware with smart mathematical tools.

Due to the given limitations in the depth (number of operations) and the width (number of qubits) of a quantum circuit, many quantum algorithms that are tested and run on existing quantum hardware are \emph{variational quantum algorithms}. 
These algorithms are based on short-depth, parameterized (variational) quantum circuits, where the parameters are trained with classical optimization tools to get a certain type of quantum state measurement.
The choice of a suitable Ansatz circuit is crucial to achieve good results. However, it is typically a-priori unknown what choice would be suitable. This issue can lead to limitations.
We would like to point out that the quantum circuit, suggested in this work, does not rely on a variational Ansatz. Instead, we provide a constructive derivation for a circuit that generates the desired statistics.

\section{Extracting Real Parts}\label{sec:real}
 The real part of any complex $z\in\mC$ can be written as $(z+\bar{z})/2$ where $\bar{z}$ denotes the complex conjugate of $z$. Similarly, $(U+U^{\dagger})/2$ extracts $\Re U$ such that it may act on the $n$-qubit state $\ket{\psi}$. To see how this can be implemented on a gate-based quantum computer, as e.g., in~\cite{LaflammeSimulatingPhysPhenom02}, consider the unitary $R = \ket{0}\bra{0} \otimes U + \ket{1}\bra{1} \otimes U^{\dagger} $.
To construct $R$, we introduce an additional, so called, auxiliary qubit. When we initialize this extra qubit $\ket{a}$ with $\ket{0}$, the joint state of the system is $\ket{0}\otimes\ket{\psi}$. 
Now, bringing the extra qubit in the superposition state $\ket{+}$ and running the circuit, we get the state $R\ (H \ket{0}\otimes\ket{\psi})$. 
Notably, the $\ket{0}$ ($\ket{1}$) component of $\ket{a}$ controls the application of $U$ ($U^\dagger$) onto $\ket{\psi}$, weighted equally by $1/\sqrt{2}$. 
Finally, another $H$-gate is applied to $\ket{a}$  and the auxiliary qubit is measured.
The action of the real part extraction is derived as follows:
$(H \otimes \In)\ R\ (H \otimes \In)\  (\ket{0}\otimes\ket{\psi})=$ 
\begin{equation*}
\nicefrac{1}{2}
\begin{pmatrix}
U+U^\dagger&U-U^\dagger\\U-U^\dagger&U+U^\dagger
\end{pmatrix}
\begin{pmatrix}
\ket{\psi}\\0
\end{pmatrix}
=
\nicefrac{1}{2}
\begin{pmatrix}
(U+U^\dagger)\ket{\psi}\\(U-U^\dagger)\ket{\psi}
\end{pmatrix}\;.
\end{equation*}
Clearly, when we measure $\ket{a}=\ket{0}$, then the output of the circuit is successful, i.e., $\ket{\psi}_{+} = \nicefrac{1}{2}(U+U^\dagger)\ket{\psi}=(\Re U)\ket{\psi}$. On the other hand, when we measure $\ket{a}=\ket{1}$, then the output of the circuit is $\ket{\psi}_{-} = \nicefrac{1}{2}(U-U^\dagger)\ket{\psi}=(\Im U)\ket{\psi}$, and thus, is incorrect. %
The implication are discussed in Section~\ref{sec:limits}

\section{Proof of Theorem \ref{thm:pmcorrectness}}\label{proof:pmcorrectness}
Without loss of generality, let $V=[n]$ and $0\leq j< 2^n$. 
When $v=1$---after the first iteration---we have to distinguish the cases $v\in\cC$ and $v\not\in\cC$. When $v\not\in\cC$, $\Phi^1_{C,\by} = I$, which satisfies the statement of the theorem due to the empty product in the definition of the sufficient statistic being $1$ and thus $(\Phi^1_{C,\by})_{j,j} = 1 = \prod_{w\in C\cap[v]} \mathbbm{1}_{\{\bx_{C\cap[v]} = \by\}}$. If instead $v\in\cC$, we have $\Phi^1_{C,\by}=(I+Z)/2$ and %
$\Phi^1_{C,\by}= (I-Z)/2$ %
for $\by_1=0$ and $\by_1=1$, respectively. 
Since $\cX_{[1]}=\{0,1\}$, the statement holds. Now, consider the induction step $v\to v+1$.
For $\Phi_{C,\by}^{v+1} = \Phi_{C,\by}^{v} \otimes A$, we have to distinguish three cases, namely: \[
A = \begin{cases}
I&\text{, if }v+1 \not\in C\\
(I-Z)/2&\text{, if }v+1 \in C \wedge \by_{v+1} = 1\\
(I+Z)/2&\text{, if }v+1 \in C \wedge \by_{v+1} = 0.
\end{cases}
\]
If $A=I$, then $v+1$ is not contained in $C$ and adding that variable does not alter the value of the sufficient statistic. Specifically, $\phi_{C\cap[v],\by}(\bx^j_{[v]})=\phi_{C\cap[v+1],\by}(\bx^{k}_{[v+1]})$ for $k\in\{2j,2j+1\}$ and all $0\leq j< 2^v$. 
Thus, $\Phi_{C,\by}^{n}\otimes I$ satisfies the statement of the theorem. 
Now, consider the case where the newly added variable $v+1$ is contained in $C$. When $\phi_{C\cap[v],\by}(\bx^j_{[v]})=0$, taking the Kronecker product with $A$ implies $\phi_{C\cap[v+1],\by}(\bx^{k}_{[v+1]})=0$ for $k\in\{2j,2j+1\}$ and all $0\leq j< 2^v$. Finally, when $\phi_{C\cap[v],\by}(\bx^j_{[v]})=1$, we either have $\phi_{C\cap[v+1],\by}(\bx^{2j}_{[v+1]})=1$ or $\phi_{C\cap[v+1],\by}(\bx^{2j+1}_{[v+1]})=1$, depending on whether $\by_{v+1}$ is $0$ or $1$. Noticing that $\phi_{C\cap[v],\by}(\bx^j_{[v]})A$ realizes the appropriate situation completes the proof.
\hfill$\blacksquare$

\section{Proof of Theorem \ref{thm:hamiltonian}}\label{proof:hamiltonian}
For the graphical model, we have $\mP_{\bt}(\bX=\bx) = \exp(\bt^\top\phi(\bx) - A(\bt))$, where $\bt^\top \phi(\bx) = \sum_{C\in\cC} \sum_{y\in\cX_C} \bt_{C,\by} \phi_C(\bx)$. According to Def.~\ref{def:pm}, the $i$-th diagonal entry of each $\Phi_{C,y}$ coincides with $\phi_{C,\by}(\bx^i)$ and hence $(H_{\bt})_{i,i}=-\bt^\top \phi(\bx^i)$. Since $H_{\bt}$ is diagonal and real-valued, $(\exp_M(-H_{\bt}))_{i,i} = \exp(\bt^\top \phi(\bx^i))$. Observing that $\log\otrace{\exp(-H_{\bt})}=A(\bt)$ completes the proof.
\hfill$\blacksquare$

\section{Proof of Theorem \ref{thm:circ}}\label{proof:circ}
Graphical models with overcomplete statistics are shift invariant. To see this, let $\bbo$ be the $d$-dimensional vector of ones and $c\in\mR$ an arbitrary constant. Now, notice that 
\begin{equation*}
\mP_{\bt+\bbo c}(\bx)=\frac{
\exp((\bt+\bbo c)^{\top} \phi(\bx))
}
{
\sum_{\bx^{'}} \exp(
{(\bt+\bbo c)}^{ \top }
\phi({\bx^{'}})
)
}
= \frac{\exp(\bt^\top\phi(\bx) + c|\cC| )}{\sum_{\bx^{'}} \exp(\bt^{\top} \phi(\bx') + c|\cC| )}
= \mP_{\bt}(\bx)\;.
\end{equation*}
Equality holds, since the vector $\phi(\bx)$ is binary with exactly $|\cC|$ ones (c.f., \eqref{eq:overstat}).

Due to shift invariance, we may subtract any fixed constant from all $\bt_j$ without altering the probability mass that corresponds to the graphical model. Hence, it is safe to assume $\bt\in\mR^d_{-}$ throughout this proof without loss of generality. 
For ease of notation, let us enumerate all $(C,\by)$-pairs from $1$ to $d$. In this notation, we have $H_{\bt} = -\sum_{j=1}^d \bt_{j} \Phi_{j}$.
Define the unitary embedding $
U_j = X \otimes (\In-\Phi_j) + Z \otimes \Phi_j$ %
where $X$, $Z$, and $I$ are the respective Pauli operators. 
Set 
\begin{equation}
    \label{eq:pointwise_polynom}
{P^y}(x) = \exp(0) + (\exp(y)-\exp(0)) x^2 = c_0 + c_2 x^2.
\end{equation}
${P^y}$ is a degree-2 polynomial %
with ${P^y}(0)=\exp(0)$ and $P^{y}(1) = \exp(y)$. 
Moreover, let $
U^j(\bgam_j) = ((\exp(i \bgam_j Z)\otimes \In) U_j)^2
$. 
We have 
\begin{equation*}
\begin{aligned}
U^j(\bgam_j)
=&\left( (\exp(i\bgam_j Z) \otimes I^{\otimes n}) U_j \right)^2\\
=&((\exp(i\bgam_j Z) \otimes I^{\otimes n}) (X\otimes(I^{\otimes n}-\Phi_j))+\\
&(\exp(i\bgam_j Z) \otimes I^{\otimes n}) (Z\otimes \Phi_j) )^2\\
=&(\exp(i\bgam_j Z) X)^2 \otimes (I^{\otimes n}-\Phi_j)^2 +\\
&(\exp(i\bgam_j Z) Z)^2 \otimes \Phi_j^2 +\\
&[(\exp(i\bgam_j Z) X) \otimes (I^{\otimes n}-\Phi_j)^2,\\&(\exp(i\bgam_j Z) Z) \otimes \Phi_j]_{+}\;,
\end{aligned}
\end{equation*}
where the commutation relations of Pauli operators and the idempotence of $\Phi_j$ are used.
By virtue of trigonometric identities and identities between Pauli matrices, we derive
\begin{equation*}
\begin{aligned}
U^j(\bgam_j)
=&(\exp(i\bgam_j Z) X)^2 \otimes (I^{\otimes n}-\Phi_j)+\\
&(\exp(i\bgam_j Z) Z)^2 \otimes \Phi_j^2\\
=&(\cos(\bgam_j)X+i\sin(\bgam_j)ZX)^2 \otimes (I^{\otimes n}-\Phi_j)+\\
&\exp(i2\bgam_j Z) \otimes \Phi_j^2\\
&(I^{\otimes n}-\Phi_j)+\exp(i2\bgam_j Z) \otimes \Phi_j^2\\
=&I \otimes (I^{\otimes n}-\Phi_j) + \exp(i2\bgam_j Z) \otimes \Phi_j^2\\
=&I^{\otimes (n+1)} + (\exp(i2\bgam_j Z)-I) \otimes \Phi_j^2\;.
\end{aligned}
\end{equation*}
Extracting the real part of $U^j(\bgam_j)$, as explained in Appendix~\ref{sec:real}, 
leads to
\begin{equation*}
\begin{aligned}
\Re U^j(\bgam_j) =& I^{\otimes n+1} + (\Re\exp(i2\bgam_j Z)-I) \otimes \Phi_j^2\\
=& I^{\otimes n+1} + \left(\cos(2\bgam_j)-1\right)I\otimes \Phi_j^2\;.
\end{aligned}
\end{equation*}
Note that the involved matrices are diagonal and hence, the result is diagonal too. Inspecting the upper left $2^n \times 2^n$ block matrix reveals that each diagonal entry has the form $1+(\cos(2\bgam_j)-1)x^2$, where $x^2$ is an diagonal entry of $\Phi_j$. Equating this expression with the polynomial $P^y$ and solving for $\bgam_j$ yields 
$\bgam_j = (1/2) \arccos(\exp(\bt_j))$ for $\bt_j\in\mR_{-}$.
We thus finally have
$
\Re U^j(\bgam(y)) = P^y(U_j)
= I \otimes \exp(\bt_j \Phi_j)
$. 
The key insights to establish the second equality above are that $\Phi_j$ is binary and idempotent, and that $P^y(x)$ from Eq.~\eqref{eq:pointwise_polynom} is constructed such that it coincides with $\exp$ on $\{0,y\}$ for $x\in\{0,1\}$. %
Note that the circuit parameters $\bgam$ are computed directly from $\bt$. %
One may think about $\bgam$ as a re-parametrization of the graphical model in the function space defined by $U^{d}(\bgam({\bbt_d}))
U^{d-1}(\bgam({\bbt_{d-1}}))
\dots U^1(\bgam({\bbt_{1}}))$.
In what follows, we set $\bbt=\nicefrac{1}{2}\ \bt$, because quantum states are defined by values proportional to the square root of the actual sampling probabilities.
All $\Phi_j$ are diagonal, which allows us to write
\begin{equation}\label{eq:realfactorization}
\begin{aligned}
\prod_{j=1}^d \Re U^j(\bgam({\bbt_j}))
=&
\begin{pmatrix} \sqrt{\exp(- H_{\bt})}&0\\0&\sqrt{\exp(- H_{\bt})}\end{pmatrix}\;.
\end{aligned}
\end{equation}
Real part extraction is in general not distributive over complex matrix products.
However, expanding $j\equiv(C,\by)$,  investigating $
\prod_{j=1}^d \Re U^j(\bgam({\bbt_j})) = \prod_{C\in\cC} \prod_{\by\in\cX_C} \Re U^{C,\by}(\bgam({\bbt_{C,\by}}))
$, 
and recalling Def.~\ref{def:pm}
reveals that the indices of diagonal entries $\neq 1$ of $U^{C,\by}(\bgam({\bbt_{C,\by}}))$ are distinct for all $\by\in\cX_C$. We may hence aggregate the real part extraction for the operators $U^{C,\by}(\bgam({\bbt_{C,\by}}))$ for all $\by\in\cX_C$ without introducing any error, e.g.,\[
\Re\begin{pmatrix}
z_1&0\\
0&1
\end{pmatrix}
\Re\begin{pmatrix}
1&0\\
0&z_2
\end{pmatrix}
=
\Re\begin{pmatrix}
z_1&0\\
0&z_2
\end{pmatrix}\;,
\] for $z_1,z_2\in\mC$. This step is significant, since each real part extraction requires an additional auxiliary qubit and decreases the success probability, as described in Sec.~\ref{sec:real}.
Invoking the real part extraction, we arrive at the final expression for the circuit\begin{align}\label{eq:qcfinal}
 \bC_{\bt} =
\prod_{C\in\cC}
H_C \otimes I^{\otimes (n+1)}\prod_{\by\in\cX_C}
\left(\ket{0}\bra{0}^{\by} \otimes U^{C,\by}(\bgam({\bbt_{C,\by}}))+\ket{1}\bra{1}^{\by} \otimes {U^{C,\by}}^{\dagger}(\bgam({\bbt_{C,\by}}))\right)\;,
\end{align}
with $\ket{z}\bra{z}^{\by}$ for $z\in\left\{0, 1\right\}$ acting on the $\by^{\text{th}}$ auxiliary qubit and $H_C=I\otimes \dots \otimes H \otimes \dots \otimes I$, where $H_C$ consists of $|\cC|$ terms and the $H$-gate is applied to the qubits that corresponds to clique $C$. 
Samples from the discrete graphical model are drawn by measuring the joint state of auxiliary and target qubits $\bC_{\bt}\ket{+}^{\otimes m}$ and discarding those samples where any of the first $|\cC|$ qubits is measured as $\ket{1}$.
The statement of the theorem follows from applying the Born rule with respect to the prepared quantum state $\bC_{\bt}\ket{+}^{\otimes m}$.
\hfill$\blacksquare$

W.l.o.g., we suppress the explicit dependence on $\bgam$ and $\bbt$ in the main part of the manuscript and write $U^{C,\by}({\bt_{C,\by}}):=U^{C,\by}(\bgam({\nicefrac{1}{2}\bt_{C,\by}}))$ to simplify the notation. 
\end{document}